\documentclass[preprint,showpacs,preprintnumbers,amsmath,amssymb]{revtex4}

\usepackage{graphicx}
\usepackage{dcolumn}
\usepackage{bm}
\usepackage{amsmath,dsfont}
\usepackage{amssymb,amsthm,amscd, amsbsy, array}

\newcommand{\half}{{\scriptstyle{\frac{1}{2}}}}
\def\2{{\half}}

\def\dAlembert{\vcenter {
    \hbox {\vrule height8pt width0.4pt depth0.0pt
           \vrule height8pt width7.2pt depth-7.6pt
           \vrule height8pt width0.4pt depth0.0pt
           \kern-8pt
           \vrule height0.4pt width8pt depth0.0pt
          \,}}}
\def\bA{{\bf A}}

\def\p{{\partial}}

\def\bbeta{{\bm{\beta}}}
\def\bgamma{{\bm{\gamma}}}
\def\bomega{{\bm{\omega}}}

\def\bb{{\bf b}}
\def\ba{{\bf a}}

\def\bv{{\bf v}}

\def\bB{{\bf B}}

\def\bnabla{\mbox{\boldmath$\nabla$}}
\def\br{{\bm{r}}}
\def\beq{\begin{equation}}
\def\eeq{\end{equation}}
\def\beqa{\begin{eqnarray}}
\def\eeqa{\end{eqnarray}}
\def\nn{\nonumber}
\def\ort{{\rm o}}
\def\Ort{{\rm O}}
\def\IR{{\mathds{R}}} 

\newcommand{\cmil}{\mathfrak{cmil}}
\def\so{{\rm so}}

\def\smallover#1/#2{\hbox{$\textstyle\frac{#1}{#2}$}} %

\begin{document}

\preprint{arXiv:0906.3594 [physics]}

\title{Non-relativistic conformal symmetries in fluid mechanics}

\author{P-M Zhang}\email{zhpm-at-impcas.ac.cn}

\author{P.~A.~Horv\'athy\footnote{On leave from 
Laboratoire de Math\'ematiques et de Physique
Th\'eorique,
 Universit\'e de Tours 
(France).}}\email{ horvathy-at-lmpt.univ-tours.fr}
\affiliation{Institute of Modern Physics, Chinese Academy of Sciences
\\
Lanzhou (China)
}

\date{\today}

\begin{abstract}
The symmetries of a free incompressible fluid span 
the Galilei group, augmented with independent dilations of space and time. When the fluid is compressible, the symmetry is
enlarged to the expanded Schr\"odinger group, which also involves, in addition, Schr\"odinger expansions.  While incompressible fluid dynamics
can be derived as an appropriate non-relativistic limit of
a conformally-invariant relativistic theory,
the recently discussed Conformal Galilei group,
obtained by contraction from the relativistic
conformal group, is \emph{not} a symmetry.
This is explained by the subtleties of 
the non-relativistic limit.
\end{abstract}


\maketitle


\section{Introduction}

Non-relativistic conformal symmetries, much studied in
recent times \cite{conformal,LSZGalconf,Gopa,Ali,MaTa,DHNC}, are two-fold.

The usual one is 
Schr\"odinger symmetry \cite{Schr,Bargmann}, highlighted by
dilations and expansion,
\beqa
D~:&
&t^*=\lambda^2t,
\quad
\br^*=\lambda\br,
\label{SchDil}
\\[6pt]
K~:&
&t^*=\Omega(t)\,{t},
\quad
\br^*=\Omega(t)\,{\br},
\quad
\Omega(t)=\frac{1}{1-\kappa t}\,,
\label{SchExpan}
\eeqa
where $\lambda>0,\,\kappa\in\IR$ \cite{Schr}. These transformations
close, with time translations $t^*=t+\epsilon$, into an
$\Ort(2,1)$ group. Note that (i) time is dilated twice
w.r.t. space, i.e. the dynamical exponent is $z=2$, while (ii)
space and time expansions share the common factor 
$\Omega(t)$. 
(iii)
Schr\"odinger symmetry typically arises for massive systems, and involves the one-parameter central extension of the Galilei group.

The second type, called ``Alt''  \cite{Hen} or Conformal Galilean (CG) Symmetry   \cite{LSZGalconf,Negro,Gopa,Ali,MaTa,DHNC,Rouhani}, is more subtle. It also has an
$\Ort(2,1)$ subgroup, generated by time translations, augmented with 
\beqa
\widetilde{D}~:&
&
t^*=\lambda\,{t},
\quad
\br^*=\lambda\,{\br},
\label{CGDil}
\\[6pt]
\widetilde{K}~:&
&
t^*=\Omega(t)\,{t},
\quad
\br^*=\Omega^2(t)\,{\br},
\label{CGExpan}
\eeqa
with the same $\Omega(t)$ as for Schr\"odinger.

The characteristic features of this second type of 
non-relativistic conformal symmetry are (i) space and time are dilated equally ($z=1$) as in a relativistic theory,
(ii) under new expansions time and space have different factors, 
$\Omega$ and  
$\Omega^2$, respectively; (iii) the CG group also contains accelerations,
\beq
\bA~:\quad
t^*=t,\quad
\br^*=\br-\frac{1}{2}\ba t^2,
\label{accel}
\eeq
where $\ba\in\IR^D$. Moreover, (iv)
this extension only allows a vanishing mass. 

Owing to masslessness, it is more 
difficult to find physical systems which exhibit this
kind of symmetry \footnote{Recent results \cite{Cherniha} indicate that, within a large class,
\emph{no} physical system can carry the CG symmetry.}.  In \cite{Bhatta,Fouxon,Gopa} it was suggested
that  \emph{incompressible fluid motion},
\begin{eqnarray}
\bnabla\cdot\bv=0,
\quad
\partial_t\bv+(\bv\cdot\bnabla)\bv=-{\bnabla P},
\label{incompcontEul}
\end{eqnarray}
where $\bv$ is the velocity field,
$P$ the pressure, and the constant density is taken to be $\rho=1$,
could be an example.

There is, however, a curious disagreement among the published statements~:

Firstly, considering the non-relativistic limit of the relativistic
Navier-Stokes equations,  
Bhattacharyya Minwalla and Wadia (BMW) \cite{Bhatta} derive the infinitesimal symmetry,
\beq\begin{array}{lll}
Dv^j&=&\big(-2t\p_t-\br\cdot\bnabla-1\big)v^j,
\\[6pt]
A_iv^j&=&-t\delta_{ij}+\half t^2\p_iv^j,
\\[6pt]
B_iv^j&=&\delta_{ij}-t\p_iv^j,
\\[6pt]
Hv^j&=&-\p_tv^j,
\\[6pt]
P_iv^j&=&-\p_iv^j,
\\[6pt]
M_{ik}v^j&=&\delta_{ij}v^k-\delta_{kj}v^i
-(x^k\p_i-x^i\p_k)v^j.
\end{array}
\label{BMWalg}
\eeq
Their  algebra
contains (i) Schr\"odinger
 dilations, $D$ in (\ref{SchDil}), (ii)  accelerations,
$A_i$, but (iii) contains no expansions.

Fouxon and Oz \cite{Fouxon} find instead that the system is
scale-invariant with respect to dilations with
any dynamical exponent $z$.  CG-expansions, $\widetilde{K}$ in
(\ref{CGExpan}), are broken, but would be
restored by a suitable modification of the system,
namely when
\begin{eqnarray}
\bnabla\cdot\bv(t,\br)&=&-3a(t),
\label{modcont}
\\[8pt]
\partial_t\bv(t,\br)+(\bv\cdot\bnabla)\bv(t,\br)&=&-{\,\bnabla P}-a(t)\bv
\label{modEuler}
\end{eqnarray}
for some function $a(t)$ of time alone.

Yet other people claim \cite{Gopa} that the system (\ref{incompcontEul}) is CG-symmetric.

At last, all these statements are in sharp
contrast with what happens for a compressible fluid, which has a mass-centrally extended Schr\"odinger symmetry \cite{Schhydro,DHNC}.

The aim of this Note is to clarify and complete
these results.

First we define~: \textit{a symmetry is a transformation
which carries a solution of the equations of motion into a
solution of these same equations}. In detail,
let us assume that $\psi$, the physical  field, belongs
to some linear space, say $H$, and 
the equation of motion is $E(\psi)=0$. Then consider a space-time
transformation $(\br,t)\to(\br^*,t^*)$,
implemented as
\beq
\psi^*(\br,t)=f(t^*,\br^*)\psi(\br^*,t^*)+g(t^*,\br^*),
\label{implem}
\eeq
[where $f$ is a linear operator acting on $\psi$ and $g\in H$ a shift], is a symmetry
if $\psi^*$ satisfies $E(\psi^*)=0$ whenever $E(\psi)=0$.
For example, $f$ is a numerical factor when $\psi$ is a
scalar, or a matrix if $\psi$ is a vector, etc.
\footnote{ 
 If $\psi$ is a complex field [a wave function],  the usual implementation is
$
\psi^*(\br,t)=f(\br^*,t^*)\psi(\br^*,t^*),
$ 
 where $f$ is a complex function.
Decomposing 
into module and phase, $\sqrt{\rho}\,e^{i\theta}$
and  $f=F e^{iG}$, changes this  into 
$$
\rho^*(\br,t)=F^2\rho(\br^*,t^*),
\qquad
\theta^*(\br,t)=\theta(\br^*,t^*)+G(\br^*,t^*).
$$
Similarly, for a vector field (like the velocity field, $\bv$), rotations are implemented through $f$, which is a rotation  matrix, etc.
}.

\section{Symmetries of the incompressible fluid equations}

The equations (\ref{incompcontEul}), which describe 
an incompressible fluid with no viscosity,
are plainly  translation and rotation symmetric.
A Galilean boost,
$
\bB~:\,
t^{*}=t,\;\br^{*}\!=\!\br+\bb t,
$
implemented as
$
\bv^{*}(t,\br)=\bv(t^{*},\br^{*})-\bb,
$ 
 also leaves the 
equations of motion invariant. 
Consistently with Ref. \cite{Fouxon}, the ``free'' system, 
$P=0$, is also invariant under dilations
with \emph{any} dynamical exponent $z$, 
\begin{equation}
D^{(z)}~:\quad
t^*=\lambda^zt,
\quad
\br^*=\lambda\,\br,
\label{zdil}
\end{equation}
Attempting to implement a $z$-dilation as
$\bv^*=\lambda^a\bv$ for a
suitable exponent $a$, we find,
$$
\p_t\bv^*+(\bv^*\cdot\bnabla)\bv^*=\lambda^{a+z}
\big\{\partial_t\bv\big\} +\lambda^{2a+1}\big\{(\bv\cdot\bnabla)\bv\big\}.
$$
The two terms scale in the same way if 
$a=z-1$, i.e., when
\beq
\bv^*(t,\br)=\lambda^{z-1}\bv(t^*,\br^*).
\label{zvdilimp}
\eeq
Then  the l.h.s. of  the Euler equation in (\ref{incompcontEul}) is multiplied by $\lambda^{2z-1}$,
\beq
\p_t\bv^*+(\bv^*\cdot\bnabla)\bv^*=\lambda^{2z-1}
\Big(\partial_t\bv+(\bv\cdot\bnabla)\bv\Big).
\label{Eulchange}
\eeq
Therefore, the system is invariant if the pressure changes as
$P^*=\lambda^{2(z-1)}P$.

The incompressibility condition is also preserved,
$\bnabla\cdot\bv^*=\lambda^{z}\bnabla\cdot\bv=0$.

Consider now general expansions of the form
\begin{equation}
K^{(\alpha)}~: \quad
t^*=\Omega(t)\,t,
\quad
\br^*=\Omega^{\alpha}(t)\,\br,
\label{genexpan}
\end{equation}
where $\alpha$ is some integer and 
try
$
\bv^{*}(t,\br)=\Omega^\delta \bv(t^{*},\br^{*})+\beta \Omega^\tau \br^{*},
$
where $\delta,\beta,\tau$ are to be determined.
Then  
\beqa
&\p_t\bv^*+(\bv^*\cdot\bnabla)\bv^*=
\Omega^{\delta+2}\Big(\p_{t^*}\bv+\Omega^{\alpha+\delta-2}(\bv\cdot\bnabla^*)\bv\Big)+
\nn
\\[6pt]
&\Omega^{\delta+1}\big(
\delta\kappa+\beta\Omega^{\alpha+\tau-1}\big)\bv
+\beta\Omega^{\tau+1}\big(\tau\kappa+\alpha\kappa
+\beta\Omega^{\tau+\alpha-1}\big)\br^*
+\Omega^{\delta+1}\big(\alpha\kappa+\beta\Omega^{\alpha+\tau-1}\big)(\br^*\cdot\bnabla^*)\bv,
\nn
\eeqa
so that we must have
$
\alpha=\delta=1,
\,
\tau=0,
\,
\beta=-\kappa.
$
Thus, \emph{the only expansion which preserves the free incompressible Euler equations is that of Schr\"odinger,} $K\equiv K^{(1)}$ in
(\ref{SchExpan}), implemented as
\beq
\bv^{*}(t,\br)=\Omega(t) 
\,\bv(t^{*},\br^{*})-\kappa \br^{*}.
\label{SchExpimp}
\eeq
This implementation multiplies the free incompressible Euler equations by the common factor $\Omega^3$.

For the incompressibility condition ${\bnabla}\cdot{\bv}=0$
 we find, however, that under a Schr\"odinger expansion,
\beq
{\bnabla}\cdot{\bv^*}=\Omega {\bnabla^*}\cdot (\Omega{\bv}-\kappa{\br^*}) =\Omega^2\, {\bnabla^*} \cdot {\bv}-D\kappa\,\Omega,
\label{incompexpbroke}
\eeq
so that \emph{the invariance w.r.t. Schr\"odinger expansions is
broken}.
The symmetry of incompressible hydrodynamics
is, therefore, the \emph{Galilei group,
augmented with arbitrary dilations of space and time}. 
 
CG expansions, $K^{(2)}\equiv\widetilde{K}$ in (\ref{CGExpan}),
 must, consequently, be broken. This is confirmed by calculating, 
\beq
\bv^{*}(t,\br)=\bv(t^{*},\br^{*})-2\kappa \Omega\br=
\bv(t^{*},\br^{*})-\frac{2\kappa\br^{*}}\Omega.
\label{CGExpimp}
\eeq
The modified incompressibility condition,
(\ref{modcont}), is correct with $D=3$ and $a(t)=2\kappa\Omega(t)$, \beq
\bnabla\cdot\bv=-6\kappa\Omega.
\label{expincomp}
\eeq
However,
\beqa
\partial_t\bv^{*}+(\bv^{*}\cdot\bnabla)\bv^{*}
=\Omega^2
\Big(\partial_{t^{*}}\bv+(\bv\cdot\bnabla^{*})\bv\Big)
-2\kappa\Omega \bv^{*}-2\kappa^2\br^{*}\nn
\\
=
\Omega^2
\Big(\partial_{t^{*}}\bv+(\bv\cdot\bnabla^{*})\bv\Big)-2\kappa\Omega\bv+2\kappa^2\Omega^2\br.
\label{expEuler}
\eeqa
The last-but-first term here would fit into the
framework (\ref{modEuler}) but the last term  breaks
 the invariance 
\footnote{Fouxon and Oz \cite{Fouxon} argue that
the CG-expansion symmetry  (\ref{CGExpan})
\emph{could be restored} by the modification
(\ref{modEuler}). This would require
the pressure to transform as
$$
P^*(t,\br)=P(t^*,\br^*)+\frac{\kappa^2}{\Omega^2}{\br^*}^2,
\label{wrongrule}
$$
which is quite an arbitrary rule which is, furthermore, \emph{inconsistent} with the one, 
(\ref{exppressure})
following from the relativistic framework.}.

For accelerations, (\ref{accel}), we get,
\beqa
\bv^*(t,\br)&=&\bv(t^*,\br^*)+\ba\, t,
\label{accelimp}
\\[6pt]
\bnabla\cdot\bv^*&=&0,\quad
\partial_t\bv^{*}+(\bv^{*}\cdot\bnabla)\bv^{*}=\Big\{\partial_{t^{*}}\bv+(\bv\cdot\bnabla^{*})
\bv\Big\}+\ba.
\label{incompaccel}
\end{eqnarray}
Accelerations can, therefore, be accommodated if the extra term is absorbed into the pressure, 
$P^*=P-\ba\cdot\br^*$, as suggested in Ref. \cite{Fouxon}.
 But do we get a symmetry~? We argue that \emph{no}, since (\ref{incompcontEul}) and 
(\ref{incompaccel})  \emph{do not describe the same system}~: while (\ref{incompcontEul}) describes fluid motion in empty space, 
(\ref{incompaccel}) describe it in a \emph{constant force field}. 
This is analogous to 
that the free fall of a projectile on Earth is different from free motion.
In our opinion, the correct interpretation is that
the transformation maps ``conformally'' one system into the other \cite{DHP2} --- just like the inertial force
can compensate terrestrial gravitation in a freely falling lift.

Both the BMW algebra (\ref{BMWalg}) and  the CGA, 
(\ref{CGDil})-(\ref{CGExpan})-(\ref{accel})
 are legitimate algebras. They are both subalgebras of the conformal Milne algebra,  $\cmil(D)$, eqn $\# (4.62)$ of Ref. \cite{DHNC},
\beq
\Big(\kappa t^2+z\lambda\, t+\epsilon\Big)\p_t+
\Big(\bomega\times\br
+2\kappa t\br+\lambda\,\br-\frac12 t^2\ba+\bbeta t+\bgamma\Big)\cdot\bnabla
\label{k0cmil}
\eeq
[where we used the obvious notation for rotations
in $D=3$].

The BMW algebra
(\ref{BMWalg}) is obtained for $z=2$ and with no expansions, $\kappa=0$, and CGA is obtained  for $z=1$ and also includes CG expansions and accelerations,  (\ref{CGExpan}) and (\ref{accel}),
respectively.
However, as explained above, 
\emph{none} of these algebras is a symmetry of the incompressible equations (\ref{incompcontEul}).  

Let us now complete (\ref{incompcontEul}) by adding
 the dissipation term $\nu\,\bnabla^2\bv$ where
$\nu$ is the shear viscosity, i.e.,
consider the incompressible  Navier-Stokes equations
\begin{eqnarray}
\bnabla\cdot\bv=0,
\quad
\partial_t\bv+(\bv\cdot\bnabla)\bv={-\bnabla P+\nu\bnabla^2\bv}\ .
\label{incompNS}
\end{eqnarray}
Ignoring space and time translations,
we only study boosts and conformal transformations.

Boosts, implemented as before,
leave the viscosity term invariant,
$\nu\bnabla^2\bv^{*}=\nu\bnabla^{*2}\bv.
$
Then for a $z$-dilation, implemented as
$\bv^{*}(t,\br)=\lambda^a\bv(t^{*},\br^{*})$, we have
\begin{eqnarray*}
\partial_t\bv^{*}+(\bv^{*}\cdot\bnabla)\bv^{*}-\nu
\bnabla^2\bv^{*}
=\lambda^{z+a}\Big\{\partial_{t^{*}}\bv\Big\}+\lambda^{2a+1}
\Big\{(\bv\cdot\bnabla^{*})\bv\Big\}-\lambda^{a+2}
\Big\{\nu\,\bnabla^{*2}\bv\Big\}.
\end{eqnarray*}
From here we infer $a=1$ and $z=2.$
Thus, for the incompressible NS flow, the dissipation term reduces dilation symmetry
to Schr\"odinger dilations only. 

The added dissipation term is also consistent with Schr\"odinger
expansions (\ref{SchExpan}), which still scale
the incompressible Euler equations by $\Omega^3$,
\beq
\partial_t\bv^{*}+(\bv^{*}\cdot \bnabla)\bv^{*}
-{\nu}\bnabla^2\bv^{*}=
\Omega
^3\Big(\partial_{t^{*}}\bv+(\bv\cdot \bnabla^{*})\bv-{\nu}{\bnabla^*}^2
\bv\Big)=0.
\eeq
Due to the non-invariance of the incompressibility condition, 
$\bnabla\cdot\bv^*\neq0$,  the
full Schr\"odinger symmetry is, nevertheless, broken to 
(Galilei) $\times$ (Schr\"odinger dilations),
cf. Table \ref{incompressibletableau}.

\begin{table}[thp]
\begin{tabular}{|l|l|l|}
\hline
\quad
&free equations
&with dissipation
\\
\hline
dilatation& dilatation with arbitrary $z$&Schr dilatation ($z=2$)
\\
\hline
expansion
&no expansion&
no expansion
\\
\hline
max symmetry
&Galilei + arbitrary dilatation&
Galilei + Schr dilatation\\
\hline
\end{tabular}
\caption{\it Symmetries 
of an incompressible fluid}
\label{incompressibletableau}
\end{table}

At last, accelerations, (\ref{accel}),
also leave invariant the incompressible NS term $(\bnabla^*)^2$, and carry the empty-space NS equations into a constant-field background equations as 
with no dissipation.

In conclusion, 

- we disagree with the claim, made in Refs.
\cite{Bhatta,Fouxon,Gopa}, that accelerations would be symmetries. The transformation (\ref{modcont})-(\ref{modEuler}) is, from our point of view, 
 \emph{not} a symmetry ; 
it is rather a transformation from one system to another one.
Moreover, 
\begin{enumerate}

\item
 The BMW algebra (\ref{BMWalg}) is \emph{incomplete} in that it
only contains  $z=2$ dilations and misses the others;

\item
Ref. \cite{Fouxon} does have all dilations, but we found  that CG expansions, $\widetilde{K}$ in (\ref{CGExpan}); don't satisfy neither our 
symmetry definition, nor
the one, in Eqns (\ref{modcont}) and (\ref{modEuler}), proposed by these authors
[except when the unjustified transformation rule
in Footnote [24] is assumed].

\item
 CGA is, thereforee, \emph{not a symmetry}, due to
the failure of CGA expansions, $\widetilde{K}$. 
 Note that the authors of 
\cite{Gopa} also miss dilatations with $z\neq1$.
It is also worth mentioning that the CGA and BMW algebras are different, contradicting  what is said in \cite{Gopa}.

\end{enumerate}

Let us insist that if our disagreement about accelerations can be considered interpretational, that about 
\emph{expansions} is fundamental. CGA expansions, $\widetilde{K}$ in (\ref{CGExpan}), are \emph{never} symmetries of fluid mechanics, in any sense. 

These questions are further discussed in Section \ref{contract} from a different point of view.

\section{Symmetries of compressible fluids}

For the sake of comparison, we now shortly discuss
compressible fluids \cite{Schhydro,DHNC} from our present point of view.
Compressible
fluid motion with no external forces is described by the Navier-Stokes equations,
\begin{eqnarray}
\partial_t\rho+\bnabla\cdot(\rho\,\bv)&=&0,
\label{continuity}
\\[8pt]{\rho}\big(
\partial_t\bv+(\bv\cdot\bnabla)\bv\big)&=&-{\,\bnabla P}+\nu\bnabla^2\bv
+\big(\zeta
+\frac{1}{3}\nu\big)\bnabla(\bnabla\cdot\bv),
\label{Eulereq}
\end{eqnarray}
where $\rho$ is the density, $\nu$ and $\zeta$ 
 the shear and the bulk viscosity, respectively \cite{LanLif,JackiwFluid}.
 When the fluid is incompressible, $\bnabla\cdot\bv=0$, the last
term disappears, and we recover the equations (\ref{incompNS}).

Let us first assume that there is no dissipation, $\nu=\zeta=0$,
and also that the motion is isentropic, $\bnabla P= \rho\bnabla  V'(\rho)$ for some function $V(\rho)$ of the density called the enthalpy \cite{JackiwFluid}.

Galilean invariance can be shown as before, completing the previous implementation (\ref{SchExpimp}) by
$\rho^*(t,\br)=\rho(t^*,\br^*)$.

The ``free'' system, $P=0$, is, again,
 scale invariant under dilations
with any dynamical exponent $z$, $D^{(z)}$ in (\ref{zdil}), implemented as in 
(\ref{zvdilimp}), completed with $\rho^*=\lambda^b\rho$. This follows from 
$$
\rho^{*}\big(\partial_t\bv^{*}+(\bv^{*}\cdot\bnabla)\bv^{*}\big) 
-\nu\bnabla^2\bv^{*}=
\lambda^b\rho\left(\lambda^{z+a}\partial
_{t^{*}}\bv+\lambda^{2a+1}(\bv\cdot\bnabla^{*})\bv\right),
$$
which still requires 
$
\bv^{*}(t,x)=\lambda^{z-1}v(t^{*},x^{*}),
$
while $b$ is left undetermined.
Given $b$, the pressure has to scale as $P^*=\lambda^{2z-2+b}P$.
In the polytropic case, $P\propto(\gamma-1)\rho^{\gamma}$, for example, this
requires 
\beq
\gamma=1+\frac{2(z-1)}{b}\ .
\label{gammaval}
\eeq
Conversely, giving $\gamma$ fixes $b$.

Next, the same proof as above shows that the free compressible Euler 
equations, (\ref{Eulereq}) with $P=0$, only allow Schr\"odinger expansions, 
$K^{(1)}\equiv K$ in (\ref{SchExpan}).
This is because the key requirement of equal scaling
of the two terms is unchanged. Completing the implementation (\ref{SchExpimp}) by $\rho^*=\Omega^\sigma\rho$, a tedious calculation shows, however, that 
choosing $\sigma=D$, i.e.,
\beq
\rho^*(t,\br)=\Omega^D\rho(t^*,\br^*),
\label{compExprho}
\eeq
cancels the unwanted term $-D\kappa\Omega$ in the incompressibility equation
(\ref{incompexpbroke}) [promoted to the continuity equation (\ref{continuity})].

The Euler equation (\ref{Eulereq}) with $\nu=\zeta=0$ scales in turn, for
$P=0$, by the factor $\Omega^{3+D}$. 
To preserve the invariance under expansions, the pressure has to scale
as $P^*=\Omega^{2+D}P$ .
In the polytropic case, this fixes the exponent as
$
\gamma=1+{2}/{D},
$
which is (\ref{gammaval}) with $z=2$ and $b=D$, and is consistent with previous results \cite{DHNC,Schhydro}.

In conclusion,  for a free compressible fluid, the symmetry is
the full \emph{expanded Schr\"odinger algebra, } 
$\widetilde{sch}$, generated by \cite{DHNC} $\#
(4.14)$,
\begin{equation}
X
=
\left(\kappa{}t^2+\mu{}t+\varepsilon\right)\frac{\partial}{\partial t}
+
\left(\bomega\times\br+\kappa{}t{}\br+\lambda{}\br+\bbeta{}t+\bgamma\right)
\cdot\bnabla,
\label{schd} 
\end{equation}
where $\bomega\in\so(D)$, $\bbeta,\bgamma\in\IR^D$, and $\kappa,\mu,\lambda,\varepsilon\in\IR$ are respectively infinitesimal rotations, boosts, spatial translations, inversions, time dilations, space dilations, and time translations.  

Do dilations and expansions combine into a  closed algebra~?
Commuting Schr\"odinger expansions with $z$-dilations,  we have,
\beq
[D^{(z)},K]=zK,
\quad
[D^{(z)},H]=-zH,
\quad
[K,H]=D^{(2)}\equiv D,
\label{tildeO21}
\eeq
 so that an $\ort(2,1)$ is only obtained when $z=2$,
 when $D^{(2)}=D\equiv D^{(sch)}$. For this value of
 $z$ (\ref{schd}) reduces to the
 Schr\"odinger algebra $sch$ i.e., the $\mu=2\lambda$  subalgebra
of the expanded Schr\"odinger algebra $\widetilde{sch}$.
Choosing  the
potential  to be consistent with $z=2$, 
yields the  Schr\"odinger symmetry in the polytropic case,
 cf. \cite{DHNC,Schhydro}.
 
For $z\neq2$ i.e. $\mu\neq2\lambda$, we only get a closed subalgebra when expansion are eliminated. Then (\ref{schd}) reduces to the Galilei algebra,
augmented with $z$-dilations \cite{DHNC}.

Furthermore, implementing accelerations as in
(\ref{accelimp}) completed with
$\rho(t,\br)=\rho(t^*,\br^*)$, 
changes the l.h.s. of the free Euler equation into
\beq
\rho^{*}\big(\partial_t\bv^{*}+(\bv^{*}\cdot\bnabla)
\bv^{*}\big)=
\rho\big(\partial_{t^{*}}\bv+
(\bv\cdot\bnabla^*)\bv\big)+\ba\rho,
\label{compaccel}
\eeq
which are the compressible Euler equations in a constant external field.
Eqn. (\ref{compaccel}) trivially generalizes
(\ref{incompaccel}).
The $\rho$-derivative terms in continuity equation cancel  and (\ref{continuity}) is acceleration-invariant.

Let us now restore the (manifestly Galilei invariant) viscosity term,
\beq
\nu\bnabla^2\bv
+\big(\zeta
+\frac 13\nu \big) \bnabla (\bnabla\cdot\bv).
\label{viscosity}
\eeq 
Both terms in (\ref{viscosity}) behaves
nicely~: instead of reducing the dilations to
$z=2$ as in the incompressible case, they allow any dynamical exponent,
fixing the scaling of $\rho$ to $b=2-z$. The pressure would have to scale as $P^*=\lambda^zP$.
Expansions are, however, broken,
leaving us with a (Galilei)$\times$(arbitrary dilation)
symmetry \cite{Schhydro}.
(In the incompressible case, freezing $\rho$
to a constant value would require $b=0$, 
yielding, once again, $z=2$.). 
The situation is summarized in Table \ref{compressibletableau}.

\begin{table}[thp]
\begin{tabular}{|l|l|l|}
\hline
\quad
&free equations
&with dissipation
\\
\hline
dilatation& $z$ arbitrary & $z$ arbitrary
\\
\hline
expansion
&Schr expansion ($\alpha=1$)
&no expansion
\\
\hline
max symmetry
&expanded Schr\"odinger $\widetilde{Sch}$ &
Galilei + arbitrary dilatation
\\
\hline
\end{tabular}
\caption{\it Symmetries of a compressible fluid}
\label{compressibletableau}
\end{table}

\section{How is relativistic conformal symmetry lost~?}\label{contract}

To get further insight, let us review
the derivation of the non-relativistic system
\cite{F-O,Fouxon}.
The starting point is 
to write the equations of 
 relativistic conformal hydrodynamics as conservation of the energy-momentum tensor \cite{F-O,Fouxon},
\begin{equation}
\partial_\nu T^{\mu \nu }=0,\;\;\;T_{\ \mu}^\mu =0.
\label{tmunucons}
\end{equation}
With 
$T_{\mu\nu}=aT^4(\eta_{\mu\nu}+4u_\mu u_\nu),$
where $T$ is the temperature  and $u_\mu $ is the four-velocity
of the fluid, this becomes 
\begin{equation}
u^\alpha \partial_\alpha \xi =-\frac{1}{3}\partial_\nu u^\nu
,\;\;\;\;\;u^\alpha\partial_\alpha u^\mu =-\partial^\mu\xi+\frac
{1}{3}u^\mu \partial_\nu u^\nu,  
\label{Req2}
\end{equation}
where $\xi=\ln T$.

To derive the non-relativistic limit, we
express our equations in terms of the  $3$-velocity, \goodbreak
\beqa
\big(1-(v/c)^2\big)\left[\delta_{ik}%
-\frac{2}{3c^2}\frac{v_iv_k}{\left[1-\smallover1/3(v/c)^2\right]}\right]\partial_k\xi + \qquad\qquad\quad
\nn
\\[4pt]
\frac{1}{c^2}\left(\partial_tv_i+v_j\partial_jv_i
-\frac{1}{3}\left(\frac{1}{\left[1-\smallover1/3(v/c)^2\right]}\right)%
v_i\,\partial_kv_k\right)
=0,
\label{xieq}
\\[16pt] 
\partial_t\xi +\frac{2}{3}\left(\frac{1}{\left[1-\smallover1/3(v/c)^2\right]}\right)\,v_i\,\partial_i\xi +
\frac{1}{3}\left(\frac{1}{\left[1-\smallover1/3(v/c)^2\right]}\right)
\partial_iv_i=0. 
\label{Req3}
\eeqa

 Keeping the leading terms only as $c\to\infty$ would yield
\beqa
\partial_i\xi=0,
\label{xiconst}
\\
\partial_t\xi+\frac{1}{3}\bnabla\cdot\bv=0,
\label{nonincomp}
\eeqa
The first equation here 
 requires that the temperature be homogenous over the whole space, and only depend on time. 
The second equation 
 is a generalization of the incompressibility equations $\bnabla\cdot\bv=0$
[to which it reduces when the temperature is constant].
No Euler equation is obtained at this order, though.
This ``simple non-relativistic limit''
is \emph{unsatisfactory} therefore, since it does \emph{not} yield the correct equations of non-relativistic hydrodynamics.
 
Interestingly, if the temperature homogeneous, (\ref{xiconst}), the second line in (\ref{xieq}) yields 
\beq
\partial_t\bv+(\bv\cdot\bnabla)\bv
-\frac{1}{3}\,\bv\,(\bnabla\cdot\bv)=0,
\qquad
\label{plessEuler}
\eeq
which \emph{is} an Euler-type equation
(\ref{Eulereq}) with $\rho=1$, no viscosity ($\nu=0$)  
no pressure ($P=0$) but with an extra term,  
$-\frac{1}{3}\,\bv\,(\bnabla\cdot\bv)=
\p_t\xi\,\bv$, completed with  (\ref{xiconst}) and
 (\ref{nonincomp}). 
For  constant temperature $T=T_0$, we would get in particular the free incompressible Euler equations (\ref{incompcontEul}). Let us also stress that (\ref{plessEuler}) comes not from the leading, 
only from the $c^{-2}$ term.

 Fouxon and Oz propose, instead, a \emph{different} kind of NR limit
reminiscent of the ``Jackiw-Nair limit''  encountered before in non-commutative mechanics \cite{JaNa}. Their clue is \emph{not} to keep 
 the term $\xi=\ln T$ finite, but put rather
\beq
P=c^2\xi=c^2\ln T
\label{pressure}
\eeq
and require that $P$, identified with the pressure, remains finite
as $c\to\infty$. Doing so allows them to recover the
\emph{incompressible Euler equations with pressure},
\beq
\bnabla\cdot\bv=0,
\qquad
\partial_t\bv+(\bv\cdot\bnabla)\bv+\bnabla P=0.
\label{incEuler}
\eeq

So far so good. But what about symmetries ?

Firstly, it is an easy matter to prove \cite{Fouxon} that the relativistic conformal
group $\Ort(4,2)$ 
is a symmetry of the relativistic system (\ref{tmunucons})
[alias (\ref{Req2})]. This is true, in particular, for [relativistic] 
\emph{special conformal  conformal transformations},
\begin{equation}
\Phi^\mu (x,b)=\frac{x^\mu+b^\mu x^2}{1+2b\cdot x+b^2x^2}\ ,  
\label{Rsct1}
\end{equation}
[where $x$ and $b$ are four-vectors],
implemented on the fields as
\begin{equation}
u_\mu(x,b)=(1+2b\cdot x+b^2x^2)\left(\partial_\mu\Phi^\alpha\right)
u_\alpha(\Phi),\qquad
T(x,b)=\frac{T(\Phi)}{1+2b\cdot x+b^2x^2}\,.
\label{Rsct2}
\end{equation}

It is also true that the contraction  $c\to\infty$ of the relativistic
conformal group is the conformal Galilei Group \cite{Barut,LSZGalconf}. A special conformal transformation becomes, in particular, a CG expansion, $\widetilde{K}$ in (\ref{CGExpan}) with $\kappa=b^0/c$, and an acceleration, $\bA$ in (\ref{accel}) with $a^i=b^i/c$.

$\bullet$  For the expansions we get, for example, 
\beqa
\bv^{*}(t,\br)=\bv(t^{*},\br^{*})+2\frac{\ba t-\kappa\br}{1-\kappa t},  
\qquad
T^{*}(t,\br)=\frac{T(t^{*},\br^{*})}{(1-\kappa t)^2}\ .
\label{NR3}
\eeqa
Note that the implementation on the velocity
is consistent with (\ref{CGExpimp}).

Then a straightforward calculation shows that 
\begin{eqnarray*}
\partial_t\xi^{*}+\frac{1}{3}\bnabla\cdot\bv^{*} 
=
\Omega^2\left(\partial_{t^{*}}\xi+\frac{1}{3}\bnabla^{*}\cdot\bv\right),
\qquad
\bnabla\xi^{*}=\Omega^2\left(\bnabla^{*}\xi\right),
\end{eqnarray*}
so that the leading-in-$c$ [but physically uninteresting] 
equations (\ref{nonincomp})-(\ref{xiconst}) \emph{are invariant}
under a CG expansion $\widetilde{K}$ in 
(\ref{CGExpan}).

The next ($c^{-2}$) order term yields, however, the free 
Euler-type equations (\ref{plessEuler}), whose ``Euler part'' is, as seen
before, \emph{invariant} under
\emph{Schr\"odinger} but not under CG expansions. The new term,
$\bv(\bnabla\cdot\bv)$, breaks, however 
both expansions.
 The full system 
(\ref{xiconst})-(\ref{nonincomp})-(\ref{plessEuler}), obtained by $c^{-2}$ truncation,  has therefore \emph{no} expansion symmetry.

But what about the incompressible Euler system (\ref{incEuler}) derived
 by the ``Jackiw-Nair type'' NR limit (\ref{pressure})~? Using (\ref{NR3}) the pressure transforms as
\beq
P^{*}(t,\br)=P(t^{*},\br^{*})-2c^2\ln (1-\kappa t).
\label{exppressure}
\eeq
The extra term here (which drops out, however, from $\bnabla P$) diverges as  $c^2\to\infty$, so finiteness of $P$ already rules out expansions. Furthermore,
\beqa
\bnabla\cdot\bv^{*}=\Omega^2\bnabla^{*}\cdot\bv-
\frac{6\kappa}{1-\kappa t},&
\label{expincompbis}
\\[12pt]
\partial_t\bv^{*}+(\bv^{*}\cdot\bnabla)\bv^{*}-\frac
{1}{3}\,\bv^{*}(\bnabla\cdot\bv^{*})+\bnabla P^{*}=
&
\nn
\\[8pt]
\Omega^2\left(\partial_{t^{*}}\bv+(\bv\cdot\bnabla^*)\bv-\frac{1}{3}\bv(\bnabla^{*}\cdot\bv)
+\bnabla^{*}P\right)&
-\,\displaystyle\frac{2\kappa^2}{(1-\kappa t)^2}%
\br+ \displaystyle{\frac{1}{3}\frac{2\kappa}{(1-\kappa t)^3}\br(\bnabla^{*}\cdot\bv})
\label{expEulerbis}
\eeqa
If  $P$ must remain finite,
then $\bnabla^*\cdot\bv=0$ and the last term drops out. Then
 (\ref{expincompbis}) and (\ref{expEulerbis}) reduce
precisely to (\ref{expincomp}) and (\ref{expEuler}),
leading to the same conclusion as before~: \emph{expansions are broken and can not be restored as
suggested in} \cite{Fouxon}.

Thus,  while the ``simple limit" (\ref{xiconst})-(\ref{nonincomp}) would have the CG-expansion symmetry obtained by
contraction, the tricky ``JN-type'' limit breaks it.

Now we can explain also the other peculiarities.

$\bullet$ As said already,
the relativistic system (\ref{Req2}) is symmetric
also under the space part of special conformal transformations, (\ref{Rsct1})-(\ref{Rsct2});
their contraction is the acceleration (\ref{accel}), 
which acts
on NR space-time and fields as in (\ref{accelimp}).
A similar calculation as above [whose details are omitted] yields, furthermore, that accelerations

\begin{enumerate}

\item 
are symmetries for the leading-in-$c$
system (\ref{xiconst})-(\ref{nonincomp});

\item leave invariant the incompressibility condition
but shift the Euler equation, as in (\ref{incompaccel})~;

\end{enumerate}

$\bullet$ A peculiar feature of incompressible hydrodynamics is that  not only CG (or Schr\"odinger), but \emph{all dilations} act as symmetries. Remarkably, this can also be explained from
considering the R $\to$ NR transition.
It is enough to discuss time dilations alone,
\beq
D^{\infty}\quad:\quad
t^*=\lambda t,\qquad
\br^*=\br,
\label{timedil}
\eeq
since all values of the dynamical exponent can
be obtained by combining $D^{\infty}$ with 
$\widetilde{D}$. Firstly, for the non-relativistic
system, we check that time dilation, (\ref{timedil}),
\begin{enumerate}
\item  implemented as
\begin{equation}
v^{*}=\lambda^pv,\;\;\;\;\xi ^{*}=\lambda ^{p-1}\xi   \label{timeD1}
\end{equation}
is a symmetry for the leading-in-$c$ system
(\ref{xiconst})-(\ref{nonincomp}) for any $p$;

\item
leave the full incompressible system (\ref{incEuler})  invariant, when
\begin{equation}
\bv^{*}(t,\mathbf{r})=\lambda\bv(t^{*},\br^{*}),
\qquad
P^{*}(t,\mathbf{r})=\lambda^2P(t^{*},\mathbf{r}^{*})
\label{timeD2}
\end{equation}

\item
 is \emph{not} a
symmetry for the relativistic system.
For the implementation (\ref{timeD2}), for example, \begin{eqnarray*}
&&\partial_t\xi^{*}+\frac{2c^2}{3c^2-\bv^{*2}}
(\bv^{*}\cdot\bnabla)\xi^{*}+%
\frac{c^2}{3c^2-\bv^{*2}}\bnabla\cdot\bv^{*} =
\\[8pt]
&&\lambda^{3}\left\{\partial_{t^{*}}\xi+\frac{2c^2}{%
3c^2-\lambda^{2}\bv^2}\bv\cdot\bnabla^{*}\xi\right\}+\lambda\frac{c^2}{%
3c^2-\lambda^{2}\bv^2}\bnabla^{*}\cdot\bv
\end{eqnarray*}
and
\begin{eqnarray*}
&&\partial_tv_i^{*}+(\bv^{*}\cdot\bnabla)v_i^{*}
-\frac{c^2-\bv^{*2}}{3c^2-\bv^{*2}}v_i^{*}
(\bnabla\cdot\bv^{*})+\left(c^2-\bv^{*2}\right)
\left[\delta_{ik}-\frac{2v_i^{*}v_k^{*}}{3c^2-\bv^{*2}}\right]\partial
_k\xi^{*} =
\\[8pt]
&&\lambda^{2}\left\{\partial_{t^{*}}v_i+(\bv\cdot\bnabla^{*})v_i
-\frac{c^2-\lambda^{2}\bv^2}{3c^2-\lambda^{2}\bv^2}v_i(\bnabla^{*}\cdot\bv)
+(c^2-\lambda^{2}\bv^2)\left[\delta_{ik}-\lambda^{2}\frac{%
2v_iv_k}{3c^2-\lambda^{2}\bv^2}\right]\partial_{k^{*}}\xi\right\}
\end{eqnarray*}
which is obviously not a symmetry if $\lambda\neq1$.
If $c\to\infty$ so that $P=c^2\xi$ remains finite, however, then all \emph{symmetry breaking terms
drop out}, and both equations scale homogeneously.
In other words, \emph{the relativistic ``no-symmetry'' (\ref{timedil})
becomes a non-relativistic symmetry}.

\end{enumerate}

\section{Conclusion}

In this paper, we have carried out a systematic study of the
conformal symmetries of non-relativistic fluids. 
 The conclusion is that the system admits 
various Schr\"odinger-type,  but no CGA-type symmetries, completing and partly
contradicting recently publicized  statements. 
In the compressible case, the new freedom of scaling the density as
in (\ref{compExprho}) allows us to overcome the 
``rigidity'' of the density and
to restore the symmetry w.r.t. \emph{Schr\"odinger expansions}, 
$K$ in (\ref{SchExpan}), which had been broken by the incompressibility  condition. 

Interesting insight can be gained when the non-relativistic systems are derived
from relativistic conformally invariant hydrodynamics. Then the leading-in-$c$ order
system does carry the CG symmetry, obtained
by contraction from the relativistic conformal group.
This system has, however, limited physical interest;
and incompressible hydrodynamics is derived
by another, more subtle limit \cite{F-O,Fouxon}, which \emph{owing to mixing different $c$-powers}, does \emph{not}
carry the CG symmetry. The general dilation symmetry
of non-relativistic hydrodynamics
is also explained from this point of view.

Our definition of a symmetry was based on the \emph{equations of motion} alone.
For a Lagrangian system however,
a symmetry can also be defined as a \emph{transformation
which changes the Lagrangian by a mere  surface term}.
This definition is clearly stronger as it implies the first one, but not vice versa~: if for example, the Lagrangian is multiplied by a constant factor then the equations of motion are preserved. 
 Note that it is only the second type of symmetries which implies, through Noether's theorem, conserved quantities. We call it, therefore, a
\emph{Noetherian symmetry}.

A compressible fluid with no dissipation can be derived from a Lagrangian \cite{JackiwFluid,Schhydro,DHNC}. The approaches based on the Lagrangian 
and on the field equations, respectively, lead to identical conclusions in the free case, but to different ones
 in the presence of a polytropic potential. In the  Lagrangian approach,
 $z$ is not more arbitrary, but fixed by the polytropic exponent. 
 
In the compressible case, the Hamiltonian structure
induced by the Lagrangian \cite{Schhydro}
could be used to provide another argument against the
CGA. One has indeed \cite{Schhydro}
\beq
\big\{\hbox{(boost)}_i,\hbox{(momentum)}_j\big\}=\hbox{(mass)}\,\delta_{ij},
\qquad
\hbox{(mass)}=\int d^3\br\,\rho>0,
\label{1centrext}
\eeq
showing that compressible fluids realize the 
\emph{one-parameter (mass) central extension}
of the Galilei group. But CGA is only consistent
with (mass)$=0$ \cite{LSZGalconf} and can not be, therefore, a symmetry.

Putting $\rho=\rho_0$ in the compressible Lagrangian
makes the system singular, making the Lagrangian approach problematic;
Hamiltonian structure should be determined using a reduction \cite{MaWe}.

\begin{acknowledgments}
P.A.H and P-M.Z are indebted to the \textit{
Institute of Modern Physics} of the Lanzhou branch of
the Chinese Academy of Sciences and to the \textit{Laboratoire de Math\'ematiques et de Physique Th\'eorique} of Tours University for hospitality, respectively.
We would like to thank C. Duval and P. Stichel for correspondence.
\end{acknowledgments}
\goodbreak


\end{document}